\documentclass[a4paper,11pt]{article}

\usepackage{jinstpub} 

\usepackage{lineno}
\usepackage{textgreek}
\usepackage{subfig}
\usepackage{url}

\title{Teaching quantum communications through a hands-on laboratory}

\author[a,b,1]{Alberto Sebastián-Lombraña,\note{Corresponding author.}}
\author[b,d]{Laura Ortiz}
\author[b,c]{Juan P. Brito}
\author[a,b]{Jaime Sáez de Buruaga}
\author[a,b]{Rafael J. Vicente}
\author[a,b]{Ruben B. Mendez}
\author[a,b]{Rafael Artiñano}
\author[b,c]{Vicente Martín}

\affiliation[a]{Universidad Politécnica de Madrid,\\Madrid, Spain}
\affiliation[b]{Center for Computational Simulation,\\Madrid, Spain}
\affiliation[c]{Dept. LSIIS, ETSIinf, Universidad Politécnica de Madrid,\\Madrid, Spain}
\affiliation[d]{Dept. DATSI, ETSIinf, Universidad Politécnica de Madrid,\\Madrid, Spain}

\emailAdd{aj.sebastian@upm.es}

\abstract{Teaching quantum communication is a challenging task when involving different technical and engineering backgrounds. The use of an approach that exploits the knowledge of these profiles, as well as other technological resources available for demonstrations or exercises, enhances this teaching. This paper presents as an example the "Quantum Communications Lab" that took place at the 6th INFIERI Summer School in 2021. In this lab, the access to the Madrid Quantum Communication Infrastructure (MadQCI) was an important resource available.}


\proceeding{International Summer School on Intelligent Signal Processing for Frontier Research and Industry\\
University Autonoma de Madrid, UAM, Madrid, Spain\\
August 23--September 4 2021
}

\begin{document}
\maketitle
\flushbottom

Quantum technologies have spread across multiple disciplines as a result of the current significant technological breakthrough~\cite{second}. Therefore, it is essential to find new approaches to adapt these new concepts to any discipline. In this work, a hands-on workshop in quantum communication oriented to research instrumentation technicians is presented.

The field of quantum computing has experienced an intense dissemination so far. However, in the specific topic of quantum communication there are not many guidelines to teach its basis. In a simple sense quantum communication deals with the transmission of the information within a quantum state between remote locations~\cite{quantum-communication}. In practice this field also involves a heterogeneous set of techniques and technologies that exploit quantum mechanics and quantum information theory to support or assist telecommunications~\cite{quantum-technologies}.

The European competence framework for quantum technologies~\cite{competence-framework} brings together four broad topics: quantum cryptography, quantum networks, infrastructure for quantum communication and hardware for quantum communications. There are examples gathered in the academical literature to teach specific fundamental techniques such as the BB84 or E91 quantum key distribution (QKD) protocols or entanglement-based repeating~\cite{undergraduate-program, incorporating-quantum}. Other examples of explaining quantum communications, such as~\cite{principles}, strongly focus on mathematical principles. Finally, there are some challenges in bringing quantum technologies to new fields;~\cite{spooky} references some examples from~\cite{grinbaum-2017, vermaas-2017, roberson-et-al, roberson}, such as focusing more in the use of quantum technologies or presenting the advantages of them in general.

Adapting quantum communications curriculum to specific profiles is a challenging task. Teaching material should be accessible and connected with the background of the students, so that it can be approached in a natural and progressive way. Quantum technologies address currently to engineering profiles, so developing a training course to address them is therefore an opportunity to both broaden the vision about quantum communication and attract these profiles. 

Research instrumentation technicians' background in signal processing or electrotechnics is ideally suited to approach quantum information processing technologies. Thus, both lecture-based and experiential-based teaching methods and resources can be adapted to them.

This script describes an approach for teaching quantum communication to research instrumentation technicians that was tested in a workshop in the 6th INFIERI Summer School, namely INtelligent signal processing for FrontIEr Research and Industry. This event focuses on teaching about cutting-edge instrumentation and signal processing as a key element of it. In that edition, it took place the course “Quantum Communications Lab” and, as a result, its intended target audience was young engineers and physicists who worked using sophisticated instrumentation daily. It was therefore a challenge to find a common teaching framework that was both motivating and accessible to these diverse profiles. 

In section~\ref{sec:tea-app}, a teaching approach to quantum communication is presented. Then, in section~\ref{sec:hand-wor} it is explained the abovementioned “Quantum Communications Lab” as an example of that teaching approach. Finally, some conclusions and the future work is compiled in section~\ref{sec:les-fut}.

\section{Teaching approach to quantum communication}
\label{sec:tea-app}

To involve a target audience, it is required an approach which enables adapting the teaching methods to its specific background. In this case, the goals are making quantum communication accessible to technical profiles while connecting with their own background.

A multidisciplinary approach allows generating accurate teaching resources but that transcend the field of mathematics and physics. In this regard, the teaching resources may show quantum communication as the natural transition of a classical communication that includes the new quantum-based and quantum-assisted techniques and technologies. And, also, this approach allows widening the view of quantum communication to many other domain-specific techniques and technologies of the students' background.

Research instrumentation technicians have different educational and professional backgrounds, including engineering, computer science and physics. It is in itself a multidisciplinary community, although they share experience and knowledge about some domain-specific fields. This is the case of the information processing technologies. They can be adopted to include quantum communication concepts, since it includes a broad set of technologies ranging from general purpose digital computers to domain-specific high-precision sensors. Also, this way, quantum communications field may be widen for adopting domain-specific techniques and technologies such as delivering metrology or synchrony information.

The following sections show an adaptation of quantum communication contents following that teaching approach.

\subsection{Quantum fundamentals adaptation}
\label{sec:quan-fund}

\begin{figure}[htbp]
\centering
\includegraphics[width=\textwidth]{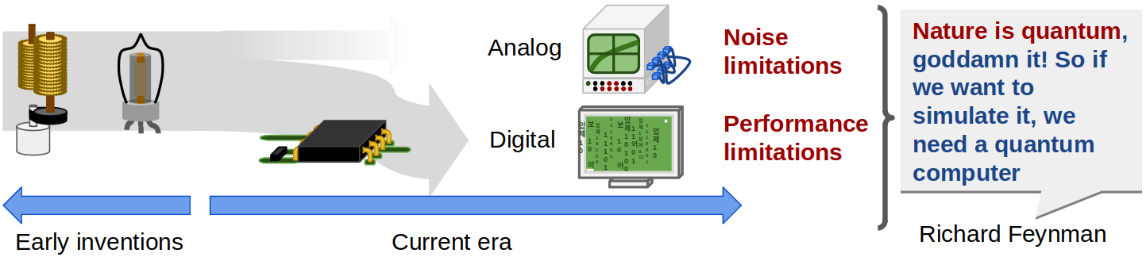}
\caption{\label{fig:evol} Example of illustration of the approach of presenting quantum technologies that we explain in the text. It is used to explain how in the middle years of the 20\textsuperscript{th} century the digital technologies prevail analogical ones due to issues with measuring uncertainty, even if the analogical approach could in theory have more computational power. This is indeed analogous to the case of quantum computing when the uncertainty principle is introduced.}
\end{figure}

Quantum technologies can be introduced as another step in the evolution of information processing technologies, as figure~\ref{fig:evol} depicts. The usual approach often uses the discrete logical or boolean bit ---~either off or on~--- for introducing the fundamental entity in quantum information, the qubit. However, the multilevel and continuous Shannon approach to information is an alternative path for that purpose. Furthermore, this perspective allows bringing students to a multidisciplinary perspective that would facilitate learning.

\begin{figure}[htbp]
\centering
\includegraphics[width=\textwidth]{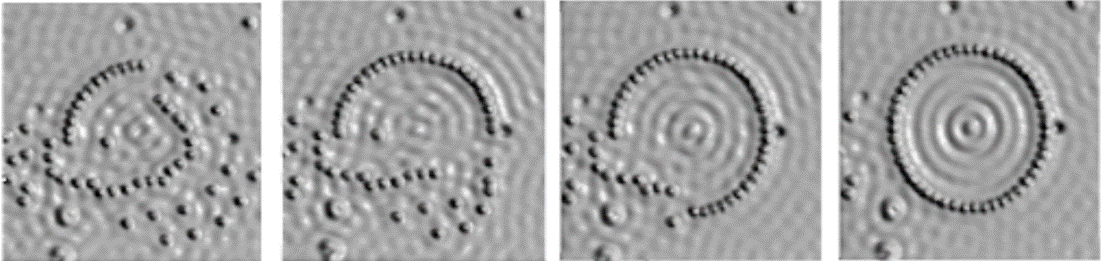}
\caption{\label{fig:ibm} “IBM's Making of a Circular Quantum Corral”, that shows various stages of moving atoms to make a circular quantum corral out of Iron atoms on a copper surface and measured with a scanning tunnelling microscope. Such an image allows showing the quantized nature of quantum systems’ properties when boundaries are set, comparing the inner and outer surface. Furthermore, it enables envisioning together with the students different technological approaches to information processing devices.}
\raggedright
Image originally created by IBM Corporation. Reproduced in accordance with IBM Copyright Permission number 30872, source \url{https://www.flickr.com/photos/21746695@N03/2104347312}.
\end{figure}

We also propose presenting quantum mechanics as a model that abstracts the complexity of any quantum physical system, e.g., electrons or photons, as an unique entity. Indeed, some specificities about quantum mechanics may be stated: the uncertainty in the measures, the superposition principle, the conservation laws and the boundary conditions. It can be done by confronting it with how classical mechanics shows the same principles. Figure~\ref{fig:ibm} shows an example of an image that helps explaining these key concepts of quantum mechanics. Also, it enables discussing with the students different possible implementations of technology based on it, such as representing, exciting and measuring several logical levels by the superposition of different standing waves or manipulating them with different excitations.

Quantum information theory may be also presented following a similar methodology. First, it may be discussed how this theory is based on resources~\cite{nielsen-chuang}, since these abstractions of quantum mechanics could embody information resources, i.e., a quantum physical system can carry, store and process information. Quantum communications therefore create, distribute, consume or delete these quantum resources, manage its life-cycle and provide access to them. And second, to move away from ground-breaking visions, quantum information may be presented with its advantages and disadvantages. Quantum technologies are often presented in the media as advantageous and even revolutionary. However, each technique and technology will use these new quantum features differently. For example, a qubit embodies more computational power, but error correction is much more complex. As a result, using qubits could be ideal in some situations but not optimal in others.

\subsection{Quantum communication techniques and technologies adaptation}
\label{sec:tec-tech}

After these fundamentals concepts, many contents could be adapted to lecture-based resources:
\begin{itemize}
\item relation between quantum information and communication, in the field of cryptography but also other domain-specific applications such as metrology or synchronization;
\item the different technical approaches, namely "quantum-assisting" and "quantum-based" systems, such as quantum key distribution or "quantum internet"~\cite{quantum-internet} respectively; 
\item some technological approaches for including smoothly these quantum technologies, as using simulators or the software-defined networking (SDN) paradigm; 
\item some specific quantum protocols, namely dense codification, teleportation or the well-known cryptographic algorithms E91 and BB84.
\end{itemize}

\begin{figure}[htbp]
\centering
\subfloat[A quantum bit representation.]{\includegraphics[width=.45\textwidth]{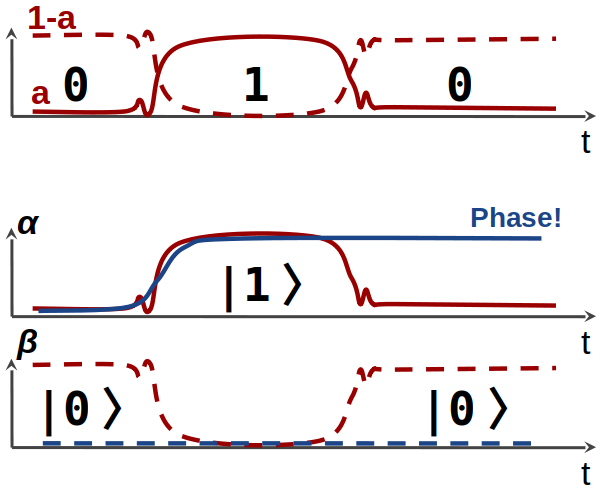}}\label{fig:qbit1}
\hfil
\subfloat[A quantum gate representation.]{\includegraphics[width=.45\textwidth]{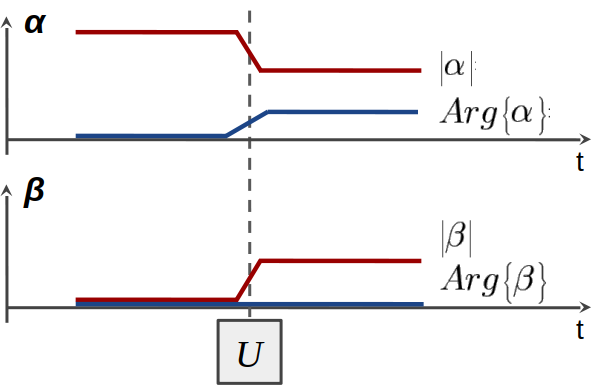}}\label{fig:qbit2}
\bigskip
\subfloat[A quantum circuit representation.]{\includegraphics[width=.90\textwidth]{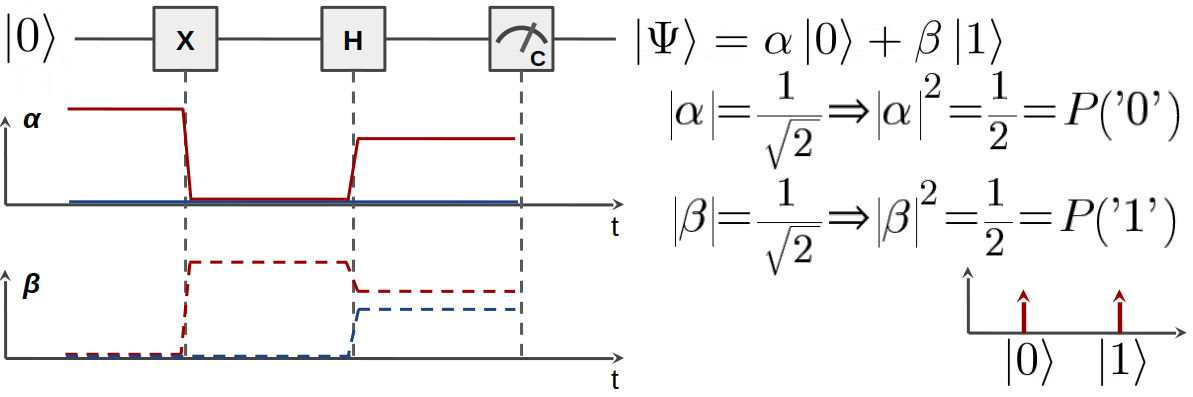}}\label{fig:qbit3}
\caption{\label{fig:qbit} This picture shows different illustrations to explain quantum information reusing the representation of information of electrotechnics or signal processing. a) depicts how the qubit can be explained as a direct evolution of the Shannon’s bit and represented as two signals superimposed. b) shows the illustration of a gate that makes a qubit evolution, in which is especially relevant showing the conservation principle. Finally, in c) is shown a whole circuit using the same formalism, and an uncertain measure process.}
\end{figure}

For this purpose, Figure~\ref{fig:qbit} represents the approach adopted for depicting the quantum bit and its temporary evolution, which seeks to resemble how electrotechnics or signal processing represent information.

\begin{figure}[htbp]
\centering
\includegraphics[width=0.9\textwidth]{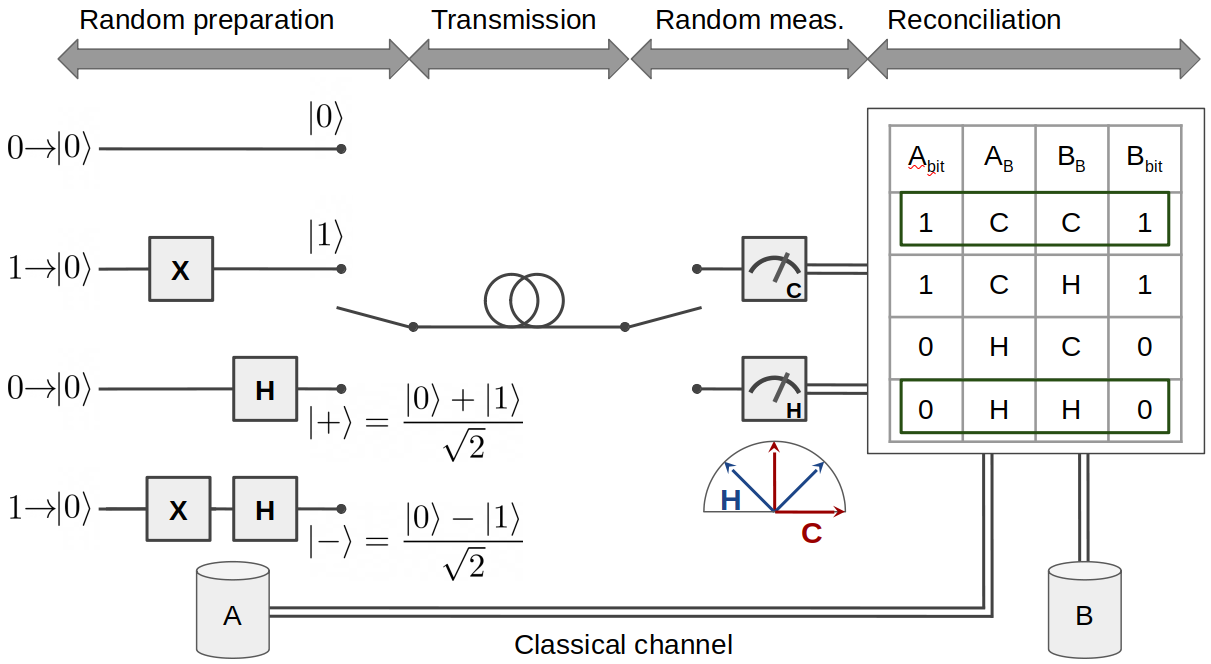}
\caption{\label{fig:bb84} The BB84 as it is explained in the approach proposed. Again, the same approach based in reusing electrotechnics’ or signal processing’s representations is used. It is for special relevance how the phase space is represented using only \textpi~radians.}
\end{figure}

Similarly, figure~\ref{fig:bb84} shows an example of the BB84 explanation, which in addition to using the same circuit formalisms as the other examples, represents the phase space in an unusual way in quantum communications. Indeed, only \textpi~radians are used, instead of the usual Bloch’s 2\textpi~radians.

\subsection{Demonstration and simulation for quantum communication}
\label{sec:demo-sim}

There are real quantum communication systems that can be used for demonstration purposes. These are those groundbreaking experiments in quantum networking performed in Boston~\cite{darpa}, Viena~\cite{viena} or Tokyo~\cite{tokio} and some more current initiatives, such as the metropolitan QKD networks in Madrid~\cite{madrid_sdn_1} or Hefei-Chaohu-Wuhu~\cite{HCW}; the long-distance 2,000~km Beijing-Shanghai QKD trunk line~\cite{china_general} and the 4,600~km long QKD space segment~\cite{china_satelite}; or the Bristol~\cite{bristol} instance, which demonstrates the penetration of these technologies in specific niches such as 5G. In the case described in this work, the course attendees had the opportunity to use the Madrid Quantum Communication Infrastructure (MadQCI), as described below.

Nevertheless, given the early level of maturity of today's quantum communications systems, a very useful resource to simulate or interact with quantum technologies are code libraries. Using them enables providing examples to support the lecture-based methodology and deliver experiential-based resources such as collaborative activities or self-evaluation exercises. It also allows to try out first-hand the new quantum features they are being introduced. So providing students with this resources is particularly useful for enabling them to develop their own designs adapted to their specific domain.

There are several libraries available currently for interfacing or simulating quantum technologies. Instances of the first branch are Qiskit~\cite{qiskit}, Cirq~\cite{quantumai}, Pyquil~\cite{pyquil}, Ocean~\cite{ocean} and PennyLane~\cite{xanadu}, that enable interfacing the real quantum technological systems of IBM, Google, Rigetti, D-Wave and Xanadu respectively. It is important to note that they are also many resources available around these libraries, as well as active communities that generate enough support. About the simulating branch, some examples include Qutip~\cite{qutip} for quantum states and Simulaqron~\cite{simulaqron}, Netsquid~\cite{netsquid}, Qunetsim~\cite{QuNetSim} or Sequence~\cite{SeQUeNCe}. 

Many of these instances are programmed in Python, a widespread language in academic domains and highly probable to be known by a technical audience. In this way, it is easy to design collaborative activities to consolidate what has been learnt. Furthermore, it is cross-platform, which would allow them to be run on the students' different machines. All libraries have specificities that make them more appropriate in each case, such as whether or not registration is required, which primitives they use (the classes' scope, whether they are technological or science-based), how difficult they are, etc. In addition, the selection of the exercises or libraries may be different depending on the intention of providing them, namely, whether they are a means of evaluation, or of promoting their adoption, etc.

It is common for demonstrations and simulations to be based on cryptographic applications of quantum communications, due to the high degree of maturity of techniques such as QKD. This is why it was possible to make the quantum resources of MadQCI available to the students, as detailed in the following section.

\section{The hands-on workshop trial "Quantum Communications Lab"}
\label{sec:hand-wor}

The course was formatted as a 3-hours workshop, which was not a simple decision but had implications for both the content and the learning outcomes. Moreover, the workshop was targeted at motivated and autonomous learners, e.g., PhD students, so it was decided that its main objective was providing attendees with the necessary tools to continue their learning afterwards. This involved a short theoretical introduction, driven by the principles exposed above, to immediately move on to setting up the working environment on the students' personal computers and completing some exercises, which were both simulation-based and using the real quantum distributed key from MadQCI. 

\begin{figure}[htbp]
\centering
\includegraphics[width=0.8\textwidth]{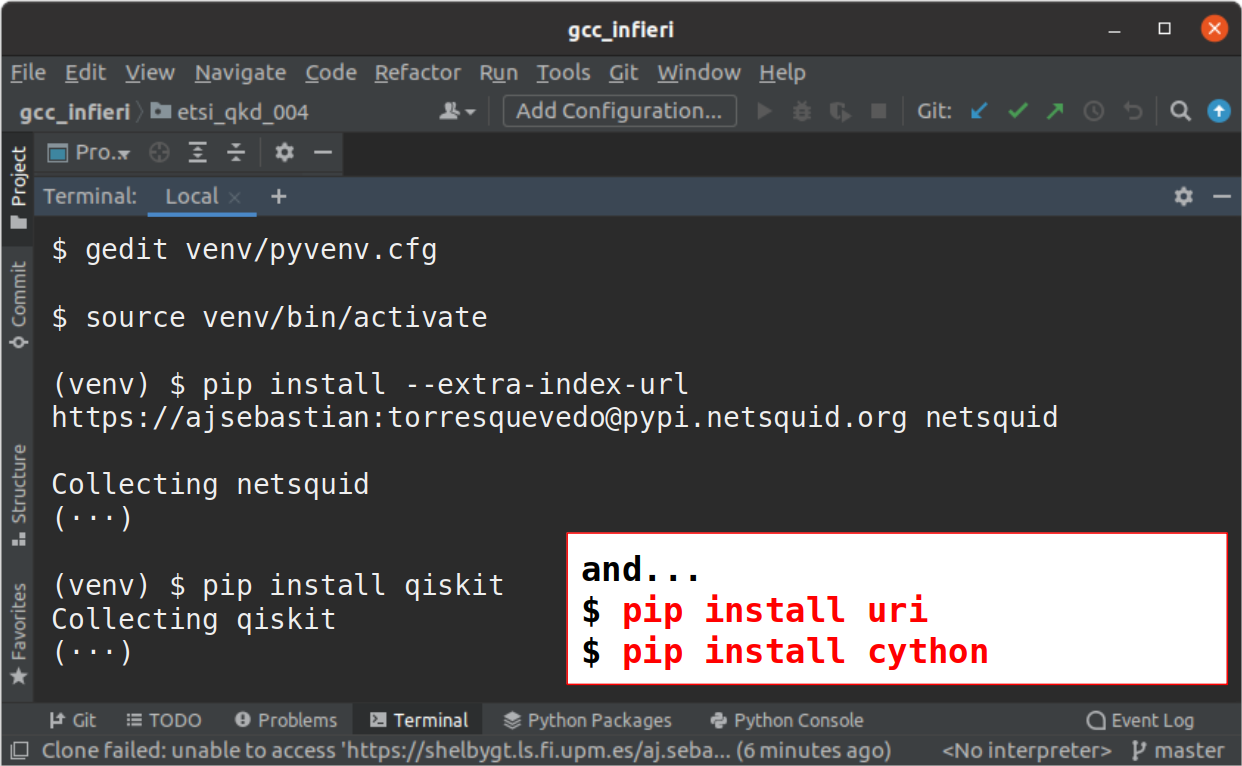}
\caption{\label{fig:snap1} Preparation of the working environment in the integrated development environment Pycharm.}
\end{figure}

Three types of teaching material were prepared for the workshop, whose purpose was mainly to provide the attendees with the necessary skills to continue their learning autonomously. The presentation began with a multidisciplinary exposition of the theoretical foundations of quantum mechanics and quantum information theory based on background of the participants in about 30 minutes. This was followed by an assisted preparation of the working environment using the participants' own computers. This preparation consisted in installing a Python distribution, learning about the Python virtual environments and installing two dependencies, Qiskit and Netsquid. The first one is a well-known library promoted by IBM and useful for simulating any quantum information system based in the circuits paradigm. Netsquid, on the contrary, was developed in Delft QuTech for simulating quantum communication systems, so it enables using communication primitives as different channel models. Figure~\ref{fig:snap1} shows an example of the media used in the workshop to illustrate this preparation. 

\begin{figure}[htbp]
\centering
\includegraphics[width=0.8\textwidth]{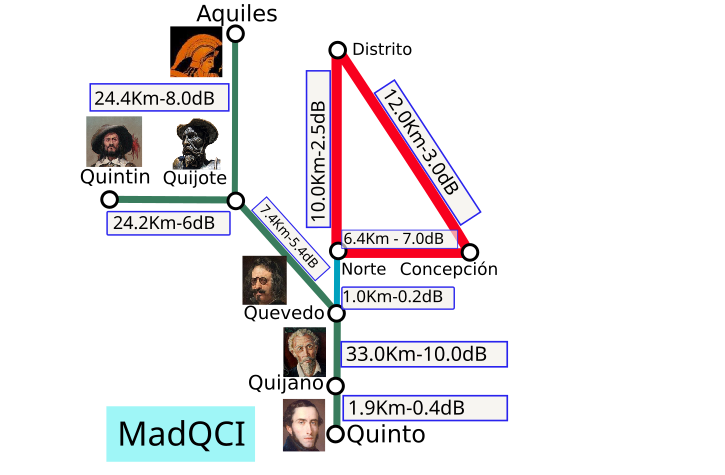}
\caption{\label{fig:madqci} Representation of the Madrid Quantum Communication Infrastructure (MadQCI). This network in Madrid is the largest test bed for quantum communication technologies in Europe and implements several standard specifications ETSI~GS~QKD.}
\end{figure}

For demonstration purposes, also the Madrid quantum network was presented. This network, named the Madrid Quantum Communication Infrastructure, MadQCI, is currently the largest test bed for quantum communication technologies in Europe and uses the software-defined approach to quantum networking for providing QKD. Figure~\ref{fig:madqci} shows a map of this infrastructure. Also, MadQCI is seamless integrated in telecommunication transport networks, both industrial and institutional domains, with multi-vendor commercial grade equipment, both classical and quantum, and implementing several standard specifications such as the ETSI~GS~QKD ones~\cite{etsi}.

\begin{figure}[htbp]
\centering
\includegraphics[width=0.8\textwidth]{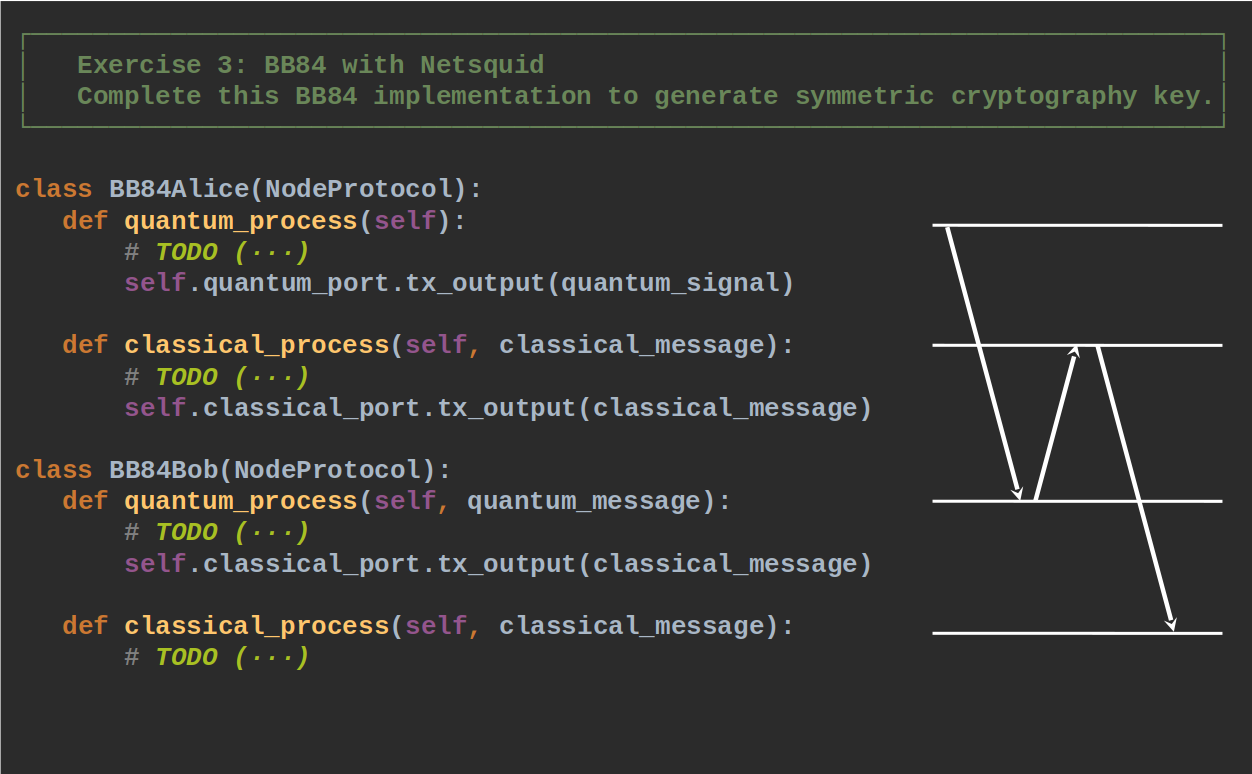}
\caption{\label{fig:snap2} Explanation of the third exercise based on Netsquid. The provided script with the exercises were, in any case, explained using Python comments.}
\end{figure}

Finally, they were provided with a single script with the proposed exercises, based on the previous preparation. The exercises were planned as an initial point of any further autonomous learning task, delivered in a single script and fully explained using Python comments. That way, as the workshop time was limited each attendee could start working when ready. However, the exercises were also explained using slides as the one shown in figure~\ref{fig:snap2}.

Six exercises were delivered for both fixing the learned contents and serve as self-evaluation:
\begin{itemize}
\item Exercise 1. This first task consisted in some minimum executions using the library Qiskit and primary was meant to detect malfunctions in the installation of the library itself. Thus, in this task, the participant could execute three different basic quantum circuits while they were encouraged to modify them as desired.
\item Exercise 2. Similarly, this one allows verifying that Netsquid is working properly. It delivers to the participant an execution that involves two quantum systems connected by a quantum channel, and it serves as the stub for the following one.
\item Exercise 3. The attendee is assisted to implement the BB84 protocol using Netsquid and, using the quantum channel models provided by this library, to introduce noise in the communication.
\item Exercise 4. In this task the attendee develops an implementation of the E91 protocol, using Qiskit and, again, with the possibility of introducing noise. The methods which trigger the execution are the same than in the previous exercise, enabling both implementations indistinguishable in the final exercise.
\item Exercise 5. The attendee can correct the induced errors in both implemented protocols using a classical error correction approach. 
\item Exercise using MadQCI. Finally, the attendee can replace the developed BB84 and E91 algorithms with a call to the MadQCI network interface to retrieve real quantum distributed key and repeat exercise 5 with this resource.
\end{itemize}
As a result, the attendee both simulates a whole quantum system, in this case a QKD one, and uses a real quantum system as is MadQCI. Most importantly, the participant ends up with a working environment ready to continue learning about quantum communications autonomously and on their own computer.

\section{Lessons learned and future work}
\label{sec:les-fut}
As quantum communication gains prominence, along with other quantum technologies, it becomes essential its teaching to professionals from many different disciplines. To this end, it is necessary adapting the content in a way that is engaging and that builds on general skills and background among engineers. And, for this, a multidisciplinary and broad teaching approach is needed for reusing the methods and domain-specific knowledge in classical technologies.

This manuscript proposes a way of teaching quantum communication to research instrumentation technicians, which share some experience and knowledge, e.g., signal processing or electrotechnics, that is ideally suited to approach quantum information processing technologies.

This way, both lecture-based and experiential-based teaching material can be adapted to involve any engineering profile in quantum communication. In this sense, the way in which the different fundamentals and techniques are represented is of great importance, as it must be adapted. For example, it is possible to explain the quantum bit with signal processing representations. In addition, it is important to use as much as possible the domain-specific knowledge that the students are familiar with, such as delivering metrology or synchrony information. 

Also, for demonstrations and simulations, both specialised code libraries and real examples of quantum communication are available. Regarding the first ones, using Python as the programming language and Qiskit and Netsquid as the code libraries for the exercises enable setting up a cross-platform environment with available resources and active on-line communities. However, in the case presented in this work, the use of a real example such as the Madrid Quantum Communication Infrastructure, MadQCI, was also of significant importance, since it allowed attendees to not only simulating an entire quantum system but using the largest European quantum communication network so far.

This teaching approach was tested in a "Quantum Communication Lab" that took place in the 6th INFIERI Summer School. In its successive editions, this event has proved to be an unrivalled framework for testing the proposed approach with high-level and highly motivated students. The course was formatted as a 3-hours hands-on workshop that involved a short theoretical introduction to immediately move on to setting up the working environments and completing some simulation-based exercises. This way, the attendees were also provided with the necessary tools to continue their learning autonomously.

The feedback from students was assessed as very positive. Given the diverse profile of the attendees, each one showed interest in different fundamentals and implications of quantum communication. This demonstrates both their engagement and the benefits of adapting the teaching material with a multidisciplinary approach that offers them a natural learning path. With a multidisciplinary teaching team, it is also possible adapting the answers and examples to the different backgrounds.

As future work, we would like to extend this approach of teaching quantum communication to other technical profiles, in order to settle a broad curriculum that involves the profiles that will actually use these quantum technologies. In addition, we want to explore other formats or ways of delivering this teaching, such as longer lectures, online experiences or written materials.

\acknowledgments
We would like to thank the organization of 6th Summer School on INtelligent signal processing for FrotlEr Research and Industry and the Universidad Autónoma de Madrid for given us the opportunity to develop this quantum communication lab. We acknowledge support from Comunidad Autonoma de Madrid, for the project QUITEMAD-CM S2018/TCS-4342; the European Union Horizon 2020 research and innovation programme under ICT grant agreement No 857156: Open European Quantum Key Distribution Testbed (OpenQKD); and also the support from Comunidad Autonoma de Madrid and Kingdom of Spain through Plan de Recuperación, Transformación y Resiliencia and European Union through NextGeneration EU founds for the founding of the project MADQuantum-CM.

\end{document}